\documentclass[a4paper,10pt]{article}

\usepackage{graphicx}
\usepackage{latexsym}
\usepackage{amsmath,amssymb} 
\usepackage{algorithm}
\usepackage{algpseudocode}
\usepackage{tikz}

\newtheorem{corollary}{Corollary}
\newtheorem{lemma}{Lemma}
\newtheorem{theorem}{Theorem}
\newenvironment{proof}{%\smallskip
\noindent{\scshape Proof:}}{\quad $\Box$\bigskip}

\algrenewcommand\algorithmicrequire{\textbf{Input:}}
\algrenewcommand\algorithmicensure{\textbf{Output:}}

\newcommand{\itemmap}{\sigma}
\newcommand{\Item}{I_\sigma}
\newcommand{\Max}{{\mathcal C}_{\mathrm{max}}}

\newcommand{\MS}{{\mathcal S}}
\newcommand{\MC}{{\mathcal C}}
\newcommand{\MB}{{\mathcal B}}

\renewcommand{\algref}[1]{Algorithm~\ref{alg:#1}}
\newcommand{\angleb}[1]{\langle#1\rangle}
\newcommand{\corref}[1]{Corollary~\ref{cor:#1}}
\newcommand{\figref}[1]{Figure~\ref{fig:#1}}
\newcommand{\tabref}[1]{Table~\ref{tab:#1}}

\newcommand{\lemref}[1]{Lemma~\ref{lem:#1}}
\newcommand{\lineref}[1]{line~\ref{line:#1}}
\newcommand{\thmref}[1]{Theorem~\ref{thm:#1}}
\newcommand{\secref}[1]{Section~\ref{sec:#1}}

\newcommand{\Ind}{\mathsf{Ind}}
\newcommand{\oraclex}[1]{\theta_{\mathrm{#1}}}
\newcommand{\cmpx}[1]{\theta_{\mathrm{#1}}}

\long\def\invis#1{}

\author{Kazuya Haraguchi\thanks{Corresponding author. E-mail: {\tt dr.kazuya.haraguchi@gmail.com}}\and Hiroshi Nagamochi}
\title{Polynomial-delay Enumeration Algorithms in Set Systems}
\date{Department of Applied Mathematics and Physics,\\ Graduate School of Informatics, Kyoto University, Japan}

\begin{document}
\maketitle
\begin{abstract}
  We consider a set system $(V, \MC\subseteq 2^V)$ on a finite set $V$
  of elements,
  where we call a set $C\in \MC$  a {\em component}. 
  We assume that two oracles $\mathrm{L}_1$ and $\mathrm{L}_2$
  are available, where  given
  two subsets $X,Y\subseteq  V$, $\mathrm{L}_1(X,Y)$ returns 
  a maximal component $C\in \MC$ with $X\subseteq C\subseteq Y$; and 
  given a set $Y\subseteq  V$, $\mathrm{L}_2(Y)$ returns 
  all maximal components $C\in \MC$ with $C\subseteq Y$. 
  Given a system $(V,\MC)$ along with a set $I$ of items and a function $\sigma:V\to 2^I$,
  a component $C\in \MC$ is called a {\em solution}
  if the set of common items in $C$ is inclusively maximal; 
  i.e., $\bigcap_{v\in C}\sigma(v)\supsetneq \bigcap_{v\in X}\sigma(v)$
  for any component $X\in\MC$ with $C\subsetneq X$.
  We prove that there exists an  algorithm  of enumerating all solutions
  (or all components) 
  in delay bounded by a polynomial with respect to
  %the input size and the running times of the oracles.
  the input size; an upper bound on the number of maximal components that is returned by $\mathrm{L}_2$;
  and the running times of the oracles.
\end{abstract}

\section{Introduction}
\label{sec:intro}

Let $V$ be  a finite set  of elements.
A {\em set system} on a set $V$ of elements is defined to be
a pair  $(V,\MC)$ of   $V$ %of elements
and 
a family $\MC\subseteq 2^V$,
where  
a set in $\MC$ is called a {\em component}.
For a subset $X\subseteq V$ in a system $(V,\MC)$, 
a component $Z\in \MC$ with $Z\subseteq X$ is called {\em  $X$-maximal} 
if  no other component $W\in \MC$ satisfies $Z\subsetneq W\subseteq X$,
and let $\Max(X)$  denote the family of all $X$-maximal components.    
For two subsets $X\subseteq  Y\subseteq V$,
let $\Max(X;Y)$ denote the family of components $C\in \Max(Y)$
such that $X\subseteq C$.  
We call a set function  $\rho$ from $2^V$ to the set $\mathbb{R}$  of reals
a {\em volume function} if 
$\rho(X)\leq \rho(Y)$ for any subsets $X\subseteq Y\subseteq V$.
A subset $X\subseteq V$ is called {\em $\rho$-positive} if $\rho(X)> 0$.
To discuss the computational complexities for solving a problem in a system,
we assume that a system $(V,\MC)$ is implicitly given as two oracles 
$\mathrm{L}_1$ and  $\mathrm{L}_2$ such that
\begin{itemize}
\item[-]
  given non-empty  subsets $X\subseteq Y\subseteq V$, 
  $\mathrm{L}_1(X,Y)$  returns a component $Z\in \Max(X;Y)$ 
    (or $\emptyset$ if no such $Z$ exists)
  in $\theta_{\mathrm{1,t}}$ time and  
  $\theta_{\mathrm{1,s}}$ space; and 
\item[-] 
  given a non-empty subset $Y\subseteq V$, 
  $\mathrm{L}_2(Y)$  returns $\Max(Y)$
  in $\theta_{\mathrm{2,t}}$ time and  
  $\theta_{\mathrm{2,s}}$ space.
\end{itemize}
We call L$_1$ and L$_2$ {\em maximal oracles}. 
Given a volume function $\rho$,
we assume that whether $\rho(X)> 0$ holds or not can be
tested in $\theta_{\rho,\mathrm{t}}$ time 
and $\theta_{\rho,\mathrm{s}}$ space. 
We also denote by $\delta(X)$ an upper bound on $|\Max(X)|$,
where we assume that $\delta$ is a non-decreasing function in the sense that
$\delta(X)\leq \delta(Y)$ holds for any subsets $X\subseteq Y\subseteq V$.

We define an {\em instance} to be a tuple $\mathcal{I}=(V,\MC,I,\sigma)$  of  
a set $V$ of $n\geq 1$ elements, a family $\MC\subseteq 2^V$, 
a set $I$ of $q\geq 1$ items and a function $\sigma:V\to 2^I$.
Let  $\mathcal{I}=(V,\MC,I,\sigma)$ be an  instance.  
The common item set $\Item(X)$ over a subset $X\subseteq V$
is defined to be  $\Item(X) = \bigcap_{v\in X}\itemmap(v)$.
A  {\em solution\/} to instance $\mathcal{I}$ is defined 
to be a component $X\in \MC$
such that  
%\begin{center}
  every component $Y\in \MC$ with $Y\supsetneq X$ satisfies  
  $\Item(Y)\subsetneq \Item(X)$. 
%\end{center} 
Let $\MS$ denote the family of all solutions to   instance  $\mathcal{I}$.
Our aim is to design an efficient algorithm for enumerating all solutions in $\MS$.  

We call  an enumeration algorithm  $\mathcal A$
\begin{itemize}
\item[-]  {\em output-polynomial}
  if the overall computation time is polynomial with respect to the input and output size;
\item[-] {\em incremental-polynomial}  
  if the computation time between the $i$-th output and 
  the $(i-1)$-st output is bounded by a polynomial with respect to
  the input size and $i$; and
\item[-] {\em polynomial-delay}  if the delay (i.e., the time between any two consecutive outputs),
  preprocessing time and postprocessing time are all bounded by a polynomial 
  with respect to the input size.
\end{itemize}

In this paper, we design an algorithm
that enumerates all solutions in  $\mathcal S$
by traversing a {\em family tree} over the solutions in  $\mathcal S$,
where the family tree is a tree structure that represents 
a parent-child relationship among solutions.
The following theorem summarizes our main result.

\begin{theorem}\label{thm:main}
  Let $\mathcal{I}=(V,\MC,I,\sigma)$ be 
  an instance on a set  system $(V,\MC)$ with a volume function $\rho:2^V\to\mathbb{R}$,
  where $n=|V|$ and $q=|I|$. 
  All $\rho$-positive solutions in $\mathcal S$ 
  to the instance  $\mathcal{I}$ can be enumerated
  in $O\big( (n+q)q\delta(V)\oraclex{1,t}+q\oraclex{2,t}+q\delta(V)\oraclex{\rho,t}+(n^2+nq)q\delta(V) \big)$
  delay and
  in $O\big(n\oraclex{1,s}+n\oraclex{2,s}+n\oraclex{\rho,s}+(n+q)n\big)$
  space. 
\end{theorem} 

The problem is motivated by enumeration of solutions
in an instance $(V,\MC,I,\sigma)$ such that $(V,\MC)$ is confluent.
We call a system $(V,\MC)$ {\em confluent} if
any tuple of components $X,Y,Z\in\MC$ 
with $Z\subseteq X\cap Y$ implies $X\cup Y\in\MC$.
For such an instance, we proposed an algorithm in \cite{HN.2022}
that enumerates all solutions such that the delay is bounded by
a polynomial with respect to the input size, an upper bound $\delta(V)$,
and the running times of oracles.

It is natural to ask whether the result in \cite{HN.2022}
is extensible to an instance with a general set system.
This paper gives an affirmative answer to the question;
even when we have no assumption on the system $(V,\MC)$
of a given instance $(V,\MC,I,\sigma)$,
there is an algorithm that enumerates all solutions 
in polynomial-delay with respect to
the input size, an upper bound $\delta(V)$, and the running times of oracles.

Technically, the scenario of this paper
is generally the same as our paper
on a confluent system~\cite{HN.2022}. 
We show that the framework in \cite{HN.2022}
can be extensible to general set systems by making 
nontrivial devices. 

The paper is organized as follows.
We describe our background in more detail
and prepare notations in \secref{prel}.
In \secref{nontransitive}, we present a polynomial-delay algorithm
that enumerates all solutions in an instance $(V,\MC,I,\sigma)$
such that $(V,\MC)$ is an arbitrary set system. 
We also show that all components are enumerable
in polynomial-delay, using the algorithm.
Finally we conclude the paper in \secref{conc}. 

%%%%%%%%%%%%%%%%%%%%%%%%%%%%%%%%%%%%%%%%%%%%%%%%%%%%%%%%%%%%
\section{Preliminaries}
\label{sec:prel}

\subsection{Background}

When $(V,\MC)$ is confluent,
the problem of enumerating solutions in  an instance $(V,\MC,I,\sigma)$
is a generalization of what we call the connector enumeration problem for a graph. 
Suppose that we are given a tuple $(G,I,\sigma)$ with
an undirected graph $G$, a set $I$ of items,
and a function $\sigma:V(G)\to 2^I$.
For a subset $X\subseteq V(G)$, 
let $G[X]$ denote the subgraph  induced from $G$ by $X$,
and $\Item(X)$ denote the common item set $\bigcap_{u\in X}\sigma(u)$.
A subset $X\subseteq V(G)$ such that $G[X]$ is connected
called a {\em connector},
if  for any vertex $v\in V(G)\setminus X$,
$G[X\cup\{v\}]$ is not connected or 
$\Item(X\cup\{v\})\subsetneq \Item(X)$; i.e., 
there is  no proper superset $Y$ of $X$ 
such that $G[Y]$ is connected and $\Item(Y)=\Item(X)$. 
The tuple $(G,I,\sigma)$
is called a graph with an item set or an attributed graph
in the literature~\cite{BMLPB.2020,ADM.2016,LSHZ.2018}.
In \cite{SS.2008,ASA.2019}, 
an attributed graph is used to represent a biological network
and a connector is regarded as a significant structure.
For this problem, 
Boley et al.~\cite{BHPW.2010} showed the first polynomial-delay algorithm. 
Sese et al.~\cite{SSF.2010} provided a fundamental enumeration algorithm,
and its parallelization is studied in \cite{OHNYS.2014,OHNYS.2016,O.2017}. 
Haraguchi et al.~\cite{HMSN2.2018} proposed 
an alternative output-polynomial algorithm based on dynamic programming.

In \cite{HN.2022}, we proposed
a polynomial-delay algorithm for 
enumerating solutions in an instance $(V,\MC,I,\sigma)$
such that $(V,\MC)$ is confluent. 
This algorithm yields polynomial-delay algorithms
for enumerating connectors in an attributed graph
and for enumerating all subgraphs with various types of connectivities
such as all $k$-edge/vertex-connected induced subgraphs
and all $k$-edge/vertex-connected spanning subgraphs
in a given undirected/directed graph for a fixed $k$.

\begin{figure}[t!]
  \centering
  %%% REV01 %%%
  %\includegraphics[width=0.8\textwidth]{inclusion_v02.eps}
  \includegraphics[width=0.7\textwidth]{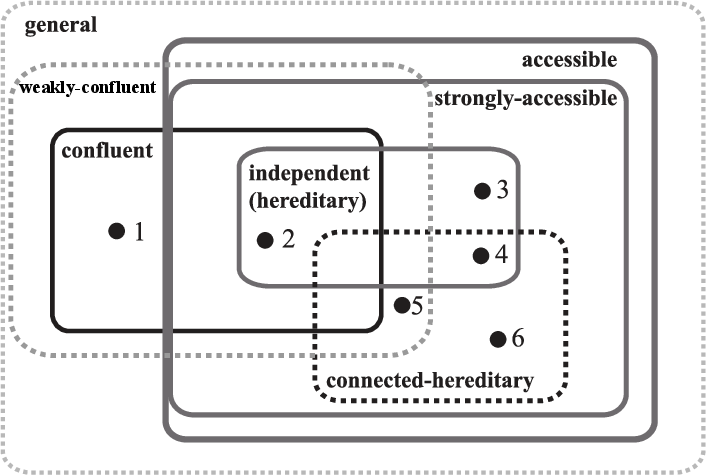}
  %\includegraphics[width=10cm]{inclusion_v02-eps-converted-to.pdf}
  %%%%%%%%%%%%%
    \begin{tabular}{ll}
    \hline
    Number & Examples\\
    \hline
    %%1 & $k$-edge/vertex-connected subgraphs, \\
    %%& spanning subgraphs (represented by edge subsets)\\
    1 & $k$-edge/vertex-connected subgraphs 
     (the empty vertex set is not contained; c.f., 5),\\
    & dominating sets, vertex covers,\\
    & spanning subgraphs (represented by edge subsets)\\
    \hline
    2 & induced subgraphs (i.e., all $2^{|V|}$ vertex subsets)\\
    \hline
    3 & independent sets, induced bipartite subgraphs, induced forests, planar graphs\\
    \hline
    4 & cliques\\
    \hline
    5 & connected induced subgraphs (the empty vertex set is contained; c.f., 1)\\
    \hline
    6 & \textsc{bc}-cliques~\cite{CGMV.2019}, connected
     $k$-plexes, induced stars, induced trees,\\
    & induced paths, connected induced bipartite subgraphs\\
    \hline
    \end{tabular}
  \caption{Venn diagram representing the relationships among 
  representative set systems; taken from Figure~2 in \cite{HN.2022}}
  \label{fig:venn}
\end{figure}

To describe the motivation of the present work,
let us review some set systems that are well-known in the literature.
\figref{venn} illustrates relationships
of some set systems that have been studied well
in the literature. 
A system $(V,\MC)$ is called 
\begin{itemize}
\item[-] {\em independent} %(or an {\em independence system})
  if, for $Y\subseteq V$, $Y\in \MC$ implies $X\in\MC$ for any $X\subseteq Y$; 
\item[-] {\em accessible}
  if, for any non-empty $Y\in\MC$,
  there is $a\in Y$ such that $Y\setminus\{a\}\in\MC$; 
\item[-] {\em strongly-accessible}
  if $\emptyset\in \MC$ and %it is accessible and,
  any pair of components $X,Y\in\MC$ with $X\subsetneq Y$
  admits an element $a\in Y\setminus X$ such that $X\cup\{a\}\in\MC$;  and
\item[-] {\em weakly-confluent}
  if, for every three components $X,Y,Z\in\MC$,
  $Z\subseteq X\cap Y$ and $Z\ne\emptyset$
  imply $X\cup Y\in\MC$. 
\end{itemize}
In the context of subgraph enumeration,
we are often asked to solve the problem as follows;
given a graph $G=(V,E)$ and a graph property $\pi$,
enumerate all subsets of $V$
that induce a subgraph that satisfies $\pi$.
%%% REV01:7 %%%
Let $\MC_\pi$ denote the set of all vertex subsets of $V$
that induce a subgraph that satisfies $\pi$.
We call the %set
system $(V,\MC_\pi)$ {\em induced by $G$ and $\pi$}. 
We call a graph property $\pi$ {\em hereditary}
if, for any $X\subseteq Y\subseteq V$,
$Y\in\MC_\pi$ implies $X\in\MC_\pi$.
We call a graph property $\pi$ {\em connected-hereditary}
if, for any $X\subseteq Y\subseteq V$,
$X\in\MC_\pi$ holds whenever 
$Y\in\MC_\pi$ and $G[X]$ is connected.
%
%%%%%%%%%%%%%%%

In \cite{HN.2022}, we assumed maximal oracles L$_1$ and L$_2$,
whereas only the membership oracle
(i.e., whether $C\in\MC$ or not is identified
for $C\subseteq V$) is assumed in studies
on the accessible system (and its subsystems). 
In the accessible system,
starting from an initial component $C\in\MC$,
we can search $\MC$ by proceeding to
$C\cup\{x\}$ for some $x\in V\setminus C$
or to $C\setminus\{y\}$ for some $y\in C$
since there is at least one component among them,
where the membership oracle identifies
whether $C\cup\{x\}$ (or $C\setminus\{y\}$)
is a component or not. 
The search strategy cannot be applied in a confluent system
since it is possible that
there is no $x\in V\setminus C$ such that $C\cup\{x\}\in\MC$
or no $y\in C$ such that $C\setminus\{y\}\in\MC$. 
This was why we introduced maximal oracles in \cite{HN.2022}.

Then it is a natural research question to ask
whether the approach assuming maximal oracles
can be extensible to more general set systems. 
The present paper shows that,
for any instance $(V,\MC,I,\sigma)$
such that $(V,\MC)$ is an arbitrary set system, 
all solutions (or all components)
can be enumerated in polynomial-delay.

To implement our framework,
it is necessary to realize maximal oracles
as a concrete and efficient algorithm,
whereas enumeration of maximal subsets
is usually non-trivial and has been a significant research issue itself.
Lawler et al.~\cite{LLK.1980} showed that,
given an independent set system,
it is generally impossible
to enumerate all maximal subsets in output-polynomial time
unless $P=NP$, but it is possible
in a special case where the {\em restricted problem} is
solvable in polynomial time.
The restricted problem is formulated as follows. 

\begin{description}
\item[\underline{Restricted problem}]
\item[Input:] A set system $(V,\MC)$,
  a component $X\in\MC$, and an element $x\in V\setminus X$.
\item[Output:] Enumerate all maximal components
  in the set system $(X\cup\{x\}, \{Y\in\MC\mid Y\subseteq X\cup\{x\}\})$. 
\end{description}

Based on Lawler et al.'s result,
complexity of enumerating maximal subsets has been analyzed for 
hereditary/connected-hereditary systems~\cite{CKS.2008,CGMV.2019}
and strongly-accessible systems~\cite{CGMV.2019}. 
\tabref{maximal} summarizes recent results
given by Conte et~al.~\cite{CGMV.2019},
where a set system is implicitly given by a membership oracle,
a subroutine for the restricted problem is assumed
to be given,
and $\cmpx{r,t}$ and $\cmpx{r,s}$ denote
the time and space complexities of the restricted problem.
As an alternative approach, Conte et al.~\cite{CMGUV.2019}
proposed proximity search,
which is a novel framework of polynomial-delay enumeration
that does not utilize the restricted problem.
It is shown to yield polynomial-delay algorithms
for various maximal subgraph enumeration problems. 

\begin{table}[t!]
  \centering
  \caption{Complexity of enumerating maximal subsets for
    various set systems $(V,\MC)$~\cite{CGMV.2019},
    where $n=|V|$ and $w=\max_{C\in\MC}|C|$}
  \label{tab:maximal}
  \begin{tabular}{llll}
    \hline
    Set system & Time && Space \\
    \hline
    Independent & $O(\textrm{poly}(n+\cmpx{r,t}))$ & (delay) & $O(\cmpx{m,s}+\cmpx{r,s})$ \\
    Connected-hereditary & $O(\textrm{poly}(n+\cmpx{r,t}))$ & (delay) & $O(\cmpx{m,s}+\cmpx{r,s})$\\
    Strongly-accessible & $O(2^w)$ & & $O(\cmpx{m,s})$ \\
    \hline
  \end{tabular}
\end{table}

%% Some previous studies deal with enumeration in these set systems.
%% Among these are
%% enumeration of maximal subsets in an independence system~\cite{JYP.1988,LLK.1980};
%% maximal induced subgraphs that satisfy a
%% given  property~\cite{CKS.2008};
%% and maximal subsets in a strongly-accessible set system~\cite{CGMV.2019}. 

%% We do not use a closure operator oracle
%% that is used in \cite{AU.2009,BHPW.2010}
%% since, without accessibility,
%% it would be hard to realize an efficient closure operator. 
%% %
%% We use oracles concerning maximal components instead. 
%% Some of previous studies (e.g., \cite{CKS.2008,LLK.1980}) utilize
%% this kind of oracle as a subroutine to solve
%% the {\em input-restricted problem}.

\subsection{Notations}

Let $\mathbb{R}$ (resp., $\mathbb{R}_+$) denote the set of reals 
(resp., non-negative reals).
For a function $f: A\to \mathbb{R}$ for a finite subset $A$ 
and a subset $B\subseteq A$,
we let $f(B)$  denote  $\sum_{a\in B}f(a)$. 

For two integers $a$ and $b$, let $[a,b]$ denote the set of integers
$i$ with $a\leq i\leq b$.
For a set $A$ with a total order $<$ over the elements in $A$,
we define a total order $\prec$ over the subsets of $A$ as follows.
For two  subsets $J,K\subseteq A$, 
we denote by $J\prec K$
if the minimum element in $(J\setminus K)\cup(K\setminus J)$ belongs to $J$.
We denote $J\preceq K$ if $J\prec K$ or $J=K$.
Note that $J\preceq K$ holds whenever $J\supseteq K$. 
Let $a_{\max}$ denote the maximum element in $A$. 
Then $J\prec K$ holds for 
$J=\{j_1,j_2,\ldots,j_{|J|}\}$, $j_1<j_2<\cdots<j_{|J|}$ and 
$K=\{k_1,k_2,\ldots,k_{|K|}\}$, $k_1<k_2<\cdots<k_{|K|}$,
if and only if 
 the sequence $(j_1,j_2,\ldots,j_{|J|},j'_{|J|+1},j'_{|J|+2},\ldots, j'_{|A|})$ 
 of length $|A|$  with $j'_{|J|+1}=j'_{|J|+2}=\cdots= j'_{|A|}=a_{\max}$
 is  lexicographically smaller than 
 the  sequence  $(k_1,k_2,\ldots,k_{|K|},k'_{|K|+1},k'_{|K|+2},\ldots, k'_{|A|})$
 of length $|A|$  with $k'_{|K|+1}=k'_{|K|+2}=\cdots=k'_{|A|}=a_{\max}$. 
Hence we see that $\preceq$ is a total order on $2^A$.

%Let $(V,\MC,I=[1,q],\sigma)$ be an instance.
Suppose that an instance $(V,\MC,I,\sigma)$ is given. 
To facilitate our aim, we introduce a total order over the items in $I$  
by representing $I$ as a set $[1,q]=\{1,2,\ldots,q\}$ of integers.
We define subsets $V_{\angleb{0}}\triangleq V$
and   $V_{\angleb{i}}\triangleq \{v\in V\mid i\in \itemmap(v)\}$
 for each item $i\in I$.
For each non-empty subset $J\subseteq I$, define subset 
$V_{\angleb{J}}\triangleq \bigcap_{i\in J}V_{\angleb{i}}=\{v\in V\mid J\subseteq \sigma(v)\}$.
For $J=\emptyset$, define $V_{\angleb{J}}\triangleq  V$.  
For each subset $X\subseteq V$, let 
$\min \Item(X)\in [0,q]$ denote the minimum item in
  $\Item(X)$, where $\min \Item(X)\triangleq 0$ for $\Item(X)=\emptyset$.
For each $i\in [0,q]$, define a family of solutions in $\MS$, 
\[\MS_i\triangleq \{X\in \MS\mid \min \Item(X)=i\}.\] 
Note that $\MS$ is a disjoint union of $\MS_i$, $i\in [0,q]$.
In Section~\ref{sec:trav},
we will design an algorithm that enumerates 
all solutions in $\MS_k$ for any specified integer $k\in [0,q]$.

For a notational convenience, 
let  $\Max(X;i)$ for each item $i\in \Item(X)$ denote 
the family $\Max(X;V_{\angleb{i}})$ of components and 
let  $\Max(X;J)$ for each subset $J\subseteq \Item(X)$
denote the  family $\Max(X;V_{\angleb{J}})$ of components. 
For a basic operation,
let us show how to test whether a given component is a solution or not.

\begin{lemma}  \label{lem:solution-test}
Let $(V,\MC,I=[1,q],\sigma)$ be an instance, 
$C$ be a component in $\MC$ and $J=\Item(C)$.
\begin{enumerate}
\item[{\rm (i)}] $C\in \MS$ if and only if 
$\{C\}=\Max(C; J)$; and 
\item[{\rm (ii)}] Whether $C$ is a solution or not can be tested
%in $O\big(?\big)$ delay and in $O\big(?\big)$ space. 
  in $O\big(\oraclex{1,t}+|C|q\big)$ delay and in $O\big(\oraclex{1,s}+|C|+q\big)$
  space. 
\end{enumerate}
\end{lemma}
\begin{proof} 
(i) Note that $C\in \MC$.
By definition, $C\not\in \MS$ if and only if
there is a component $C'\in \MS$ such that  $C\subsetneq C'$
and $\Item(C)=\Item(C')=J$, where
a maximal one of such components  $C'$ belongs to $\Max(C; J)$. %$C\in \Max(C; J)$. 
Hence if no such component $C'$ exists then 
$\Max(C; J)=\{C\}$.
Conversely, if $\Max(C; J)=\{C\}$ then no such component $C'$ exists.

(ii)
Let $Y$ be a subset such that $C\subseteq Y\subseteq V$.
We claim that
$\Max(C;Y)=\{C\}$ holds if and only if
L$_1(C,Y)$ returns the component $C$.
The necessity is obvious. For the sufficiency,
if there is $X\in\Max(C;Y)$ such that $X\ne C$,
$X$ would be a superset of $C$, contradicting the $Y$-maximality of $C$.
By (i), to identify whether $C\in \MS$ or not,
it suffices to see whether L$_1(C,J)$ returns $C$.
We can compute $J=\Item(C)$ in $O(|C|q)$ time
and in $O(q)$ space,
and can decide whether the oracle returns $C$
in $O(\oraclex{1,t}+|C|)$ time and
in $O(\oraclex{1,s}+|C|)$ space. 
%\qed
\end{proof}

%%%%%%%%%%%%%%%%%%%%%%%%%%%%%%%%%%%%%%% 
\section{Enumeration Algorithms} 
\label{sec:nontransitive}

\subsection{Family Tree}
Similarly to our work on a confluent system~\cite{HN.2022},
we use the notion of family tree to list all solutions in $\MS$ efficiently.
Our tasks to establish such an enumeration algorithm are as follows:
\begin{description}
\item[(I)] 
 Select some solutions from the set $\MS$ of solutions 
 as  the roots, called ``bases''; 
\item[(II)] Define the ``parent'' $\pi(S)\in \MS$ of 
  each non-base solution $S\in \MS$,
  where the solution $S$ is called a ``child'' of the solution $T=\pi(S)$;  
\item[(III)] Design an algorithm~A that, given a solution $S\in \MS$,
  returns its parent $\pi(S)$; and  
\item[(IV)] Design an algorithm~B that, given a solution $T\in \MS$, 
  generates a set $\mathcal{X}$ of components $X\in\MC$ such that 
  $\mathcal{X}$ contains all children of $T$. 
  We can test whether each component $X\in\mathcal{X}$ is a child of $T$
  by constructing  $\pi(X)$ by algorithm~A and checking  if  $\pi(X)$
  is equal to $T$.
\end{description}
Starting from each base, we recursively generate the children of a solution. 
The delay of the entire enumeration algorithm depends on
the time complexity of algorithms A and B, 
where $|\mathcal{X}|$ is bounded from above
by the time complexity of algorithm~B.

Roughly speaking, (II) and (III) in the present paper
are essentially different from \cite{HN.2022}
since these in \cite{HN.2022} depends on the confluency,
whereas (I) and (IV) are mostly the same as \cite{HN.2022}. 
We explain details of the four parts one by one. 

\paragraph{{\rm (I)} Defining Base}
For each integer $i\in [0,q]$, define a set of components %solutions 
$\MB_i\triangleq \{X\in \Max(V_{\angleb{i}})\mid \min \Item(X)=i\}$,
and $\MB\triangleq \bigcup_{i\in [0,q]}\MB_i$.
We call each component in $\MB$ a {\em base}. 

The following lemma is shown in \cite{HN.2022}
and used as a fundamental tool for enumerating solutions in a confluent system.
The proof does not utilize the confluency
and thus the lemma can be used in the current problem as it is.
We present the lemma and the proof just
for self-completeness of the paper. 

\begin{lemma}[Lemma 4 in \cite{HN.2022}]  \label{lem:base}
Let $(V,\MC,I=[1,q],\sigma)$ be an instance.
\begin{enumerate}
\item[{\rm (i)}] For each non-empty set $J\subseteq [1,q]$ or $J=\{0\}$, 
it holds that   $\Max(V_{\angleb{J}})\subseteq \MS$;   
\item[{\rm (ii)}] For each $i\in [0,q]$,  any solution $S\in  \MS_i$ 
  is contained in a base in $\MB_i$;  and  
\item[{\rm (iii)}] $\MS_0=\MB_0$ and  $\MS_q=\MB_q$. 
\end{enumerate}
\end{lemma}
\begin{proof} 
(i)   Let $X$ be a   component in $\Max(V_{\angleb{J}})$.
  Note that  $J\subseteq \Item(X)$ holds.
  When  $J=\{0\}$ (i.e., $V_{\angleb{J}}=V$), 
   no proper superset of $X$ is a component,   and $X$ is a solution.  
  Consider the case of $\emptyset\neq J\subseteq [1,q]$.  
  To derive a contradiction, assume that $X$ is not a solution; i.e.,
  there is a proper superset $Y$ of $X$ such that $\Item(Y)=\Item(X)$.
  Since $\emptyset\neq J\subseteq \Item(X)=\Item(Y)$, 
  we see that $V_{\angleb{J}}\supseteq Y$.
  This, however, contradicts the $V_{\angleb{J}}$-maximality of $X$. 
   This proves that $X$ is a solution. 
   
 (ii)  We prove that each solution $S\in  \MS_i$ 
 is contained in a base in $\MB_i$.
 Note that  $i=\min \Item(S)$ holds.
 By definition, it holds that $S\subseteq V_{\angleb{i}}$.
 Let $C\in  \Max(S;V_{\angleb{i}})$ be a solution. 
 Note that $\Item(S)\supseteq \Item(C)$ holds. 
 Since $i\in \Item(C)$ for $i\geq 1$
 (resp., $\Item(C)=\emptyset$ for $i=0$),
 we see that $\min \Item(S)=i=\min \Item(C)$. 
 This proves that $C$ is a base in $\MB_i$.  
 Therefore $S$ is contained in a base $C\in \MB_i$.  
 
 (iii) 
 Let $k\in \{0,q\}$. 
 We see from (i)  that $\Max(V_{\angleb{k}})\subseteq \MS$,
 which implies that  
  $\MB_k
  =\{X\in \Max(V_{\angleb{k}})\mid \min \Item(X)=k\}
  \subseteq \{X\in \MS\mid \min \Item(X)=k\} =\MS_k$.
 We prove that any solution $S\in \MS_k$ is a base in $\MB_k$.
 By (ii), there is a base $X\in \MB_k$ such that $S\subseteq X$,
 which implies that 
 $\Item(S)\supseteq \Item(X)$ and $\min\Item(S)\leq \min\Item(X)$.
 We see that  $ \Item(S)=  \Item(X)$, since 
 $\emptyset=\Item(S)\supseteq \Item(X)$ for $k=0$,
 and $q=\min\Item(S)\leq \min\Item(X)\leq q$ for $k=q$.
Hence $S\subsetneq X$ would contradict that $S$ is a solution.
Therefore 
$S=X\in  \MB_k$,
as required.  
%\qed
\end{proof}

  Lemma~\ref{lem:base}(iii) tells that all solutions in $\MS_0\cup \MS_q$
  can be found 
by calling oracle $\mathrm{L}_2(Y)$ for $Y=V_{\angleb{0}}=V$ and 
$Y=V_{\angleb{q}}$.
We concentrate on how to generate 
all solutions in $\MS_k$ for each item $k\in [1,q-1]$.

\paragraph{{\rm (II)} Defining Parent}
%
%This subsection defines the ``parent'' of a non-base solution. 
%
The definition of the parent of a non-base solution
is more complicated than our work \cite{HN.2022} on a confluent system.
When the system $(V,\MC)$ is confluent,
we define the parent $T\in\MS$ of a non-base solution $S\in\MS$
by using only the item set $\Item(T)$. 
This is possible due to the following properties
of a confluent system.
\begin{itemize}
  \item For any component $X\in\MC$ and a superset $Y\supseteq X$,
    $\Max(X;Y)$ is a singleton (Lemma~3 in \cite{HN.2022}).
  \item For two solutions $S,T\in\MS$ with $S\subseteq T$,
    $\Max(S;\Item(T))=\{T\}$ holds (Lemma~5 in \cite{HN.2022}).
\end{itemize}
However, these properties do not hold in general set systems. 
Then we need a more intricate definition of the parent.

For two subsets $X,Y\subseteq V$,  we denote 
$(\Item(X),X)\prec (\Item(Y),Y)$ 
if 
``$\Item(X)  \prec \Item(Y)$'' or
``$\Item(X) =\Item(Y)$ and $X\prec  Y$''  
and let 
$(\Item(X),X)\preceq  (\Item(Y),Y)$ mean  
$(\Item(X),X)\prec (\Item(Y),Y)$ or $X=Y$. 

Let $X\subseteq V$ be a subset such that $k=\min \Item(X)\in [1,q-1]$.
We call a solution $T\in \MS$  a {\em superset solution} of $X$ 
if  $T\supsetneq X$ and $T\in \MS_k$. 
A superset solution $T$ of $X$  is called {\em minimal} 
if no proper subset $Z\subsetneq T$ is a  superset solution of $X$.
We call a minimal superset solution $T$ of $X$ 
{\em the lex-min solution} of $X$ if 
$(\Item(T),T)\preceq  (\Item(T'), T')$ 
for all minimal superset solutions $T'$ of $X$.  
For each item  $k\in [1,q-1]$,
 we define the {\em parent} $\pi(S)$ of a non-base solution 
 $S\in \MS_k\setminus \MB_k$
  to be the lex-min solution  of $S$,
  and   define  a {\em child} of  a solution $T\in \MS_k$
  to be a non-base solution $S\in \MS_k\setminus \MB_k$ such that $\pi(S)=T$.

\paragraph{{\rm (III)} Finding the Parent (Algorithm A)}
For a non-base solution $S\in\MS\setminus\MB$,
\lemref{item_minimal} characterizes the item set $\Item(T)$
of the parent $T=\pi(S)$ and
\lemref{vertex-index-minimal} characterizes $T$
as a vertex set. 

\begin{lemma}   \label{lem:item_minimal} 
Let $(V,\MC,I=[1,q],\sigma)$ be an instance,
  $S\in \MS_k\setminus \MB_k$ be a  non-base  solution 
  for some item $k\in  [1,q-1]$, 
and  $T\in \MS_k$ denote the  lex-min solution of $S$. 
Denote $\Item(S)$ by $\{k, i_1,i_2,\ldots,i_p\}$ so that $k<i_1<i_2<\cdots<i_p$.
For each integer $j\in [1,p]$, 
 $i_j\in \Item(T)$ holds if and only if $\Max(S;J\cup\{i_j\})\neq\{S\}$ holds
for the item set   $J=\Item(T)\cap \{k, i_1,i_2,\ldots,i_{j-1}\}$.
\end{lemma}
\begin{proof}
 By Lemma~\ref{lem:base}(i) and $\min\Item(S)=k$, 
we see that $\Max(S;J\cup\{i_j\}) \subseteq \MS_k$
for any integer $j\in [1,p]$. \\
\noindent
{\bf  Case~1.} $\Max(S;J\cup\{i_j\})=\{S\}$: 
 For any subset $J'\subseteq \{i_{j+1},i_{j+2},\ldots,i_p\}$,
the family $\Max(S;J\cup\{i_j\}\cup J')$ is equal to $\{S\}$ and cannot 
contain any minimal superset solution of $S$.
This  implies that $i_j\not\in \Item(T)$.\\
\noindent
{\bf  Case~2.}  $\Max(S;J\cup\{i_j\})\neq \{S\}$:
Let $C$ be an arbitrary component in $\Max(S;J\cup\{i_j\})$.
Then $C$ is a solution by Lemma~\ref{lem:base}(i).
Observe that $k\in J\cup\{i_j\}\subseteq \Item(C)\subseteq \Item(S)$ and
$\min\Item(C)=k$, implying that $C\in \MS_k$ is a superset solution of $S$.
Then $C$ contains
a minimal superset solution $T^*\in \MS_k$ of $S$, where 
$\Item(T^*)\cap[1,i_{j-1}]=\Item(T^*)\cap \{k,i_1,i_2,\ldots,i_{j-1}\}\supseteq 
J= \Item(T)\cap \{k,i_1,i_2,\ldots,i_{j-1}\}=\Item(T)\cap[1,i_{j-1}]$ 
and $i_j \in\Item(T^*)$.
If $\Item(T^*)\cap[1,i_{j-1}]\supsetneq J$ or $i_j\not\in \Item(T)$, 
 then $\Item(T^*)\prec \Item(T)$ would hold, contradicting that $T$ is 
the lex-min solution of $S$.
Hence $\Item(T)\cap[1,i_{j-1}]=J=\Item(T^*)\cap[1,i_{j-1}]$
and    $i_j\in \Item(T)$. 
%\qed
\end{proof}

The next lemma tells us how to construct
 the parent $T=\pi(S)$ of a given solution $S$. 

\begin{lemma}  \label{lem:vertex-index-minimal} 
Let $(V,\MC,I=[1,q],\sigma)$ be an instance,
  $S\in \MS_k\setminus \MB_k$ be a  non-base  solution 
  for some item $k\in  [1,q-1]$, 
and  $T$ denote the  lex-min solution of $S$. 
Let $J=\Item(T)$.  
Let $S'$ be a set such that 
  $S\subseteq S'\subsetneq T$, 
where  $V_{\angleb{J}}\setminus S'$ is denoted by 
$\{u_i\mid i\in[1,s=|V_{\angleb{J}}|-|S'|]\}$ such that $u_1<u_2<\cdots<u_s$. 
%$\{u_i\mid i\in[1,s=n-|S'|]\}$ such that $u_1<u_2<\cdots<u_s$. 
Then: 
\begin{enumerate}
\item[{\rm (i)}]  
   $T\in \Max(S'\cup\{u \};V_{\angleb{J}})$
    for any vertex $u \in T\setminus S'$;
\item[{\rm (ii)}]  
Every component $C\in \MC$ with $S'\subsetneq C\subseteq V_{\angleb{J}}$
satisfies  $\Item(C)=J$;
\item[{\rm (iii)}]  
  There is an integer  $r\in [1,s]$   such that   
  $\Max(S'\cup\{u_j\};V_{\angleb{J}})=\emptyset$  
  for each $j\in [1,r-1]$   and 
  all components  $C\in\Max(S'\cup\{u_r\};V_{\angleb{J}})$
  satisfy  $\Item(C)=J$; 
\item[{\rm (iv)}]
For the integer $r$ in {\rm (iii)}, 
$T\cap \{u_j\mid j\in[1,r] \}=\{u_r\}$ holds; and
\item[{\rm (v)}]
For the integer $r$ in  {\rm (iii)},
if $S'\cup\{u_r\}\in \MS$ then $T=S'\cup\{u_r\}$ holds.
\end{enumerate}
\end{lemma}
\begin{proof} 
(i)  Since $S'\subsetneq T$, there exists a vertex $u \in T\setminus S'$.
  For such a vertex $u $, 
  $T$ is a component such that
  $S'\cup\{u \}\subseteq T\subseteq V_{\angleb{J}}$. 
If $T$ is not a $V_{\angleb{J}}$-maximal component, then
there would exist a component $Z\in \MC$ with 
$T\subsetneq Z\subseteq V_{\angleb{J}}$  and
$J=\Item(T)\supseteq \Item(Z)\supseteq \Item(V_{\angleb{J}})\supseteq J$, 
contradicting that $T$ is a solution.
Hence  $T\in \Max(S'\cup\{u \};V_{\angleb{J}})$
    for any vertex $u \in T\setminus S'$.
   
  (ii) 
 Let $C\in \MC$ be a component  with $S'\subseteq C\subseteq V_{\angleb{J}}$.
Note that  
 $\Item(S)\supseteq \Item(S')\supseteq  \Item(C)\supseteq 
 \Item(V_{\angleb{J}})\supseteq J=\Item(T)$ 
  and 
 $k=\min\Item(S)=\min\Item(T)$. 
Since $C$ is a component, there is a solution $S_C$ such that
$S_C\supseteq C$ and $\Item(S_C)=\Item(C)$.
Since $S\subseteq S'\subsetneq C\subseteq S_C$,
$S$ and $S_C$ are distinct solutions and there
must be a minimal superset solution $S^*_C\in \MS_k$ of $S$ such that 
$S\subsetneq S^*_C\subseteq S_C$, where we see that 
$\Item(S) \supsetneq   \Item(S^*_C)\supseteq \Item(S_C)
=\Item(C)\supseteq  \Item(T)$ and 
 $k=\min\Item(S)=\min\Item(S^*_C)  =\min\Item(T)$.
If $\Item(C)\supsetneq J$, then  
$\Item(S^*_C)\supseteq \Item(S_C)=\Item(C)\supsetneq J=\Item(T)$
  implies that 
$\Item(S^*_C)\prec \Item(T)$, contradicting that $T$ is the lex-min
solution of $S$. 
  
  (iii)  By (i), for some integer $r\in [1,s]$, 
    $T\in \Max(S'\cup\{u_r\};V_{\angleb{J}})$
    holds and some component  $C\in\Max(S'\cup\{u_r\};V_{\angleb{J}})$
  satisfies  $\Item(C)=\Item(T)=J$.
Let $r$ denote the smallest index 
such that no component $C\in\Max(S'\cup\{u_j\};V_{\angleb{J}})$
  satisfies $\Item(C)= J$  for each $j\in [1,r-1]$. 
 By (ii), for such $r$, the statement of (iii) holds. 
  
  (iv)   
   Since  no component $C\in\Max(S'\cup\{u_j\};V_{\angleb{J}})$
  satisfies $\Item(C)=J$ for all integers $j\in [1,r-1]$, 
  no  component  $T'\supsetneq S'$ such that 
$T'\cap \{u_j\mid j\in[1,r-1] \}\neq\emptyset$ 
 can be the lex-min solution $T$. 
Since   some component  $C\in\Max(S'\cup\{u_r\};V_{\angleb{J}})$
  satisfies  $\Item(C)=\Item(T)=J$, 
there is a component $T'\in \MC$ such that $u_r\in T'$ and
 $\Item(T')=J=\Item(T)$. 
 The lex-min solution $T$ satisfies $ T   \preceq  T' $  
for all minimal superset solutions $T'$ of $S$ with $\Item(T')=J$.
Therefore  $T$ must contain $u_r$.
 
(v) By (iv),  
$u_r\in T$. 
If $S'\cup\{u_r\}\in \MS$ then
$S'\cup \{u_r\}$ is a unique minimal superset solution of $S$
such that   $T\supseteq S'\cup \{u_r\}\supseteq S$, 
implying that $T=S'\cup \{u_r\}$.
%\qed
\end{proof} 

Based on Lemmas~\ref{lem:item_minimal} an
\ref{lem:vertex-index-minimal},
we describe an algorithm
to compute the parent of a given non-base solution
in \algref{parent_c}. 

\begin{lemma}   \label{lem:greedy_minimal} 
Let $(V,\MC,I=[1,q],\sigma)$ be an instance,
  $S\in \MS_k\setminus \MB_k$ be a  non-base  solution 
  for some item $k\in  [1,q-1]$.
  Then   {\sc Parent}$(S)$ in \algref{parent_c}  correctly delivers 
  the lex-min solution of $S$
  in $O\big((n+q)\oraclex{1,t}+n^2+nq\big)$
  %$O\big((n+q)\oraclex{1,t}+n\oraclex{2,t}+n^2q\delta(V_{\angleb{k}})\big)$
  time
  and in $O\big(\oraclex{1,s}+\oraclex{2,s}+n+q\big)$ space.   
\end{lemma}
\begin{proof}
  Let $T$ denote the lex-min solution of $S$. 
  The item set $J$ constructed in the first for-loop 
  (lines \ref{line:parent_former_for} to \ref{line:parent_former_endfor})
  satisfies $J=\Item(T)$ by \lemref{item_minimal}.
  The second for-loop
  (lines \ref{line:parent_latter_for} to \ref{line:parent_latter_endfor})
  picks up $u_i\in T\setminus(S\cup Z)$ by   
  \lemref{vertex-index-minimal}(iv),
  and the termination condition (\lineref{parent_if}) is
  from \lemref{vertex-index-minimal}(v). 
  
  The first for-loop  is repeated $p\le q$ times,
  where we can decide whether the condition in \lineref{parent_eq} holds
  in $O(\oraclex{1,t}+|S|)$ time and in $O(\oraclex{1,s}+|S|)$ space.
  The time and space complexities of the first for-loop
  are $O(q(\oraclex{1,t}+|S|))$ and $O(\oraclex{1,s}+|S|)$.

  We can decide the set $V_{\angleb{J}}$ in $O(nq)$ time and in $O(n+q)$ space. 
  
  The second for-loop is repeated $s\le n=|V|$ times.
  We can decide whether the condition of \lineref{parent_if} is satisfied
  by calling the oracle L$_1(S\cup Z\cup \{u_i\};V_{\angleb{J}})$,
  which takes $O(\oraclex{1,t})$ time and $O(\oraclex{1,s})$ space.
  When the condition of \lineref{parent_if} is satisfied,
  we can decide whether $S\cup Z\in\MS$ or not (\lineref{parent_sol})
  in $O(\oraclex{1,t}+|S\cup Z|q)$ time
  and in $O(\oraclex{1,s}+|S\cup Z|+q)$
  space by \lemref{solution-test}(ii).  
  The time and space complexities of the second for-loop
  are $O(n(\oraclex{1,t}+n))$
  and $O(\oraclex{1,s}+n)$.

  The overall time and space complexities are
  $O\big((n+q)\oraclex{1,t}+n^2+nq\big)$
  and $O\big(\oraclex{1,s}+\oraclex{2,s}+n+q\big)$. 
  %\qed
\end{proof} 
 
\begin{algorithm}[t!]
  \caption{{\sc Parent}$(S)$: 
  Finding the lex-min solution of a  solution $S$ }
  \label{alg:parent_c}
  \begin{algorithmic}[1]
    \Require An instance $(V,\MC,I=[1,q],\itemmap)$, 
      an item $k\in  [1,q-1]$, and
      a non-base solution $S\in \MS_k\setminus \MB_k$,
       where $k= \min \Item(S)$. 
    \Ensure
     The lex-min solution $T\in \MS_k$ of $S$. 
     \State Let $\{k,i_1,i_2,\ldots,i_p\}:=\Item(S)$, where $k<i_1<i_2<\cdots<i_p$;
     \State $J:=\{k\}$;      
     \For {{\bf each} integer $j:=1,2,\ldots,p$} 
     \label{line:parent_former_for}
         \If {$\Max(S;J\cup\{i_j\})\neq\{S\}$ } 
         \label{line:parent_eq}
            \State $J:=J\cup\{i_j\}$  
         \EndIf 
      \EndFor; \Comment{$J=\Item(T)$ holds}
      \label{line:parent_former_endfor}
     \State Let  $\{u_1,u_2,\ldots,u_s\}:= V_{\angleb{J}}\setminus S$, 
     where $u_1<u_2<\cdots<u_s$;
    \State $Z:=\emptyset$;
    \For{{\bf each} integer $i:=1,2,\dots,s$}
     \label{line:parent_latter_for} 
    \If {$\Max(S\cup Z\cup\{u_i\};V_{\angleb{J}})\neq\emptyset$} 
 %    \Comment{$\Item(C)=J$  holds for any component $C\in\Max(S\cup Z\cup\{u_i\};V_{\angleb{J}})$} 
    \label{line:parent_if}
    \State $Z:=Z\cup\{u_i\}$;
    \If{$S\cup Z\in\MS$}
    \label{line:parent_sol}
    \State Output $T:=S\cup Z$ and halt
    \EndIf
    \EndIf
    \EndFor
    \label{line:parent_latter_endfor}
  \end{algorithmic}
\end{algorithm}

\paragraph{{\rm (IV)} Generating Children (Algorithm B)}

\lemref{child_candidate} shows
how to construct a family $\mathcal{X}$ of
components for a given solution $T$ so that $\mathcal{X}$ contains
all children of $T$.
The statement of the lemma
is the same as Lemma~7 in \cite{HN.2022},
except the computational complexity in (iii),
where we provide an essentially new proof for (i)
since the proof for Lemma~7(i) in \cite{HN.2022}
utilizes the confluency and thus is not applicable to our problem. 

\begin{lemma}   \label{lem:child_candidate}  
Let $(V,\MC,I=[1,q],\sigma)$ be an instance
and   $T\in \MS_k$ be a solution for some item $k\in  [1,q-1]$.
Then:
 \begin{enumerate}
 \item[{\rm (i)}] 
   Every child $S$ of $T$   
   satisfies  $[k+1,q]\cap (\Item(S)\setminus \Item(T)) \neq\emptyset$
   and is a component in $\Max(T\cap V_{\angleb{j}})$ 
   for any item  $j\in[k+1,q]\cap (\Item(S)\setminus \Item(T))$;
 \item[{\rm (ii)}] 
The family of children $S$ of $T$ is
equal to the disjoint collection of families 
$\mathcal{C}_j =
   \{ C\in\Max(T\cap V_{\angleb{j}})\mid 
        k= \min \Item(C), C\in \MS, 
        j=\min\{i\mid i\in [k+1,q]\cap (\Item(C)\setminus \Item(T))\},
        T=${\sc Parent}$(C)\}$ over all items  
        $j\in[k+1, q]\setminus \Item(T)$; and 
 \item[{\rm (iii)}] 
  The set of all children of $T$ can be constructed in 
  $O\big((n+q)q\delta(T)\oraclex{1,t}+q\oraclex{2,t}+(n^2+nq)q\delta(T)\big)$ time and
   $O(\oraclex{1,s}+\oraclex{2,s}+n+q)$ space. 
 \end{enumerate}
\end{lemma}
\begin{proof} 
(i) 
Note that $[0,k]\cap \Item(S)=[0,k]\cap \Item(T)=\{k\}$ since $S,T\in \MS_k$. 
Since $S\subseteq T$ are both solutions,
$\Item(S)\supsetneq \Item(T)$.
Hence 
$[k+1,q]\cap (\Item(S)\setminus \Item(T)) \neq\emptyset$.
  Let $j$ be an arbitrary item in $[k+1,q]\cap (\Item(S)\setminus \Item(T))$.
  We see $S\subseteq T\cap V_{\angleb{j}}$ since
  $S\subseteq T$ and $j\in\Item(S)$.
  To show that $S$ is a component in $\Max(T\cap V_{\angleb{j}})$,
  suppose that there is a component $C\in\MC$ such that
  $S\subsetneq C\in\Max(T\cap V_{\angleb{j}})$.
  Since $\Item(S)\supseteq\Item(C)\supseteq\Item(T\cap V_{\angleb{j}})\supsetneq\Item(T)$
  and $\min\Item(S)=\min\Item(T)=k$,
  we see that $\min\Item(C)=k$.
  Then $C$ should not be a solution since otherwise
  it would be a superset solution of $S$ such that $S\subsetneq C\subsetneq T$,
  contradicting that $T$ is a minimal superset solution of $S$.
  Since $C$ is not a solution but a component,
  there is a solution $C'$ such that $C'\supsetneq C$ and $\Item(C')=\Item(C)\subseteq\{ k\}$. 
  Hence $C'\in\MS_k$. 
  Such a solution $C'$ contains a minimal superset solution $C''$ of $S$
  such that $C'\supseteq C''\supsetneq S$ and $\Item(C')\subseteq\Item(C'')\subsetneq\Item(S)$. 
  Then we have $\Item(S)\supsetneq\Item(C'')\supseteq\Item(C')=\Item(C)\supsetneq\Item(T)$,
  and thus $C''\prec T$ holds, 
  which contradicts that $T$ is the lex-min solution of $S$.
  Therefore, such $C$ does not exist, implying that $S\in\Max(T\cap V_{\angleb{j}})$.
\invis{
  Let $j$ be an arbitrary item  in $[k+1,q]\cap (\Item(S)\setminus \Item(T))$,
  it holds that $S\subseteq V_{\angleb{j}}$ and $S\subseteq T$. 
  There is no other component $C\in \MC$ such that 
  $S\subsetneq C\subsetneq T$, since otherwise
  $\Item(S)\supseteq \Item(C)\supseteq \Item(T)$ and 
  $k=\min \Item(S)=\min \Item(T)$ would imply $ \min \Item(C)= k$
  and $C\subseteq T\cap V_{\angleb{j}}$,
  contradicting that $C$ is   a   $(T\cap V_{\angleb{j}})$-maximal component
  in $\Max(T\cap V_{\angleb{j}})$.
  Hence $S\in\Max(S;T\cap V_{\angleb{j}})$
  for any item $j\in [k+1,q]\cap (\Item(S)\setminus \Item(T))$.
}

 (ii)  
 By (i), the family $\mathcal{S}_T$ of children of $T$ is 
 contained in the family of 
  $(T\cap V_{\angleb{j}})$-maximal components $C\in\MS$
  over all items $j\in [k+1,q]\cap  \Item(T)$.
Hence $\mathcal{S}_T
    =\cup_{j\in [k+1,q]\cap \Item(T)}\{C\in \Max(T\cap V_{\angleb{j}})
    \mid  C\in\MS, T=${\sc Parent}$(C)\}$.
Note that 
  if a subset $S\subseteq V$ is a child of $T$, then 
 $k= \min \Item(S)$, $C\in\MS$ and 
 $S\in \Max(T\cap V_{\angleb{j}})$ for all items 
$j\in [k+1,q]\cap (\Item(S)\setminus \Item(T))$.
Hence we see that  $\mathcal{S}_T$ is
equal to the disjoint collection of families 
$\mathcal{C}_j =
   \{ C\in\Max(T\cap V_{\angleb{j}})\mid 
        k= \min \Item(C), C\in\MS, 
        j=\min\{i\mid i\in [k+1,q]\cap (\Item(C)\setminus \Item(T))\},
        T=${\sc Parent}$(C)\}$ over all items  
        $j\in[k+1, q]\setminus \Item(T)$.   
        
        (iii)
        We show an algorithm to generate all children of $T\in\MS_k$
        in \algref{children}. The correctness directly follows from (ii). 
        The outer for-loop (lines \ref{line:children_outer_begin}
        to \ref{line:children_outer_end}) is repeated at most $q$ times.
        Computing $\Max(T\cap V_{\angleb{j}})$ in \lineref{children_l2}
        can be done in $\oraclex{2,t}$ time and in $\oraclex{2,s}$ space.
        For each $C\in\Max(T\cap V_{\angleb{j}})$,
        the complexity of deciding 
        whether $C$ satisfies the condition in \lineref{children_if} or not
        is dominated by {\sc Parent$(C)$}.
        Let $\tau$ denote the time complexity of {\sc Parent$(C)$}. 
        The time complexity of the entire algorithm is
        \[
        O\big(q(\oraclex{2,t}+\delta(T)\tau)\big)
        =O\big((n+q)q\delta(T)\oraclex{1,t}+q\oraclex{2,t}+(n^2+nq)q\delta(T)\big);
        \]
        and the space complexity is $O(\oraclex{1,s}+\oraclex{2,s}+n+q)$,
        where the computational complexities of {\sc Parent$(C)$}
        are from \lemref{greedy_minimal}. 
        %\qed
\end{proof}

\begin{algorithm}[t!]
 \caption{{\sc Children}$(T,k)$: Generating all children}\label{alg:children}
\begin{algorithmic}[1]
\Require  An instance $(V,\MC,I,\itemmap)$, an item $k\in   [1,q-1]$ and   
a solution $T\in \MS_k$. 
\Ensure All children of $T$, each of which is output whenever it is generated.
\For{{\bf each} item  $j\in[k+1, q]\setminus \Item(T)$}
\label{line:children_outer_begin}
\State Compute $\Max(T\cap V_{\angleb{j}})$;
\label{line:children_l2}
    \For{{\bf each} component $C\in\Max(T\cap V_{\angleb{j}})$} 
    \label{line:children_inner_begin}
      \If {$k= \min \Item(C)$, 
         $C\in \MS$, 
        $j=\min\{i\mid i\in [k+1,q]\cap (\Item(C)\setminus \Item(T))\}$ \par 
          \hskip\algorithmicindent and $T=${\sc Parent}$(C)$ 
          %  (i.e., $C$ is a child of $T$)  
          }
      \label{line:children_if}
                   \State Output $C$ as one of the children of $T$ 
              \EndIf    
    \EndFor
    \label{line:children_inner_end}
    \EndFor 
\label{line:children_outer_end}
  \end{algorithmic}
\end{algorithm}

%%%%%%%%%%%%%%%%%%%%%%%%%%%%%%%%%%%%%%%%%%%%%% 
\subsection{Enumerating Solutions}
\label{sec:trav}

Now we have done with preparation
of a family-tree based enumeration algorithm.
The remaining scenario is almost the same as
our work on a confluent system~\cite{HN.2022}.

We describe an entire algorithm for enumerating
solutions in $\MS_k$ %with
for a given integer $k\in [0,q]$. 
We first compute the component set $\Max(V_{\angleb{k}})$. 
We next compute the family $\MB_k~(\subseteq \Max(V_{\angleb{k}}))$ of bases
by testing whether $k=\min \Item(T)$ or not for each component $T\in \Max(V_{\angleb{k}})$. 
%where $T\in \MB_k\subseteq \MS_k$. %$\MB_k\subseteq \MS_k$.
When $k=0$ or $q$, we are done with $\MB_k=\MS_k$ 
by Lemma~\ref{lem:base}(iii). 
Let  $k\in [1,q-1]$.
Suppose that we are given a solution $T\in \MS_k$.
We find all the children of $T$ by
{\sc Children}$(T,k)$ in  Algorithm~\ref{alg:children}. 
By applying Algorithm~\ref{alg:children}
to a newly found child recursively,
we can find all solutions in $\MS_k$.

Assume that a  volume function   $\rho: 2^V\to \mathbb{R}$ 
is given. 
An algorithm that enumerates all  $\rho$-positive solutions in $\MS_k$
is described in Algorithms~\ref{alg:enumalg} and
\ref{alg:enumdesc}.
For the solution output,
we employ the {\em alternative output method\/}~\cite{U.2003}
to make the delay faster; see lines 5 to 10 in
\algref{enumdesc}.
In this method, for a given solution $T\in\MS_k$,
we output the children of $T$ after (resp., before)
generating all descendants
when the depth of the recursive call to $T$
is an even (resp., odd) integer.
By this, we can avoid traversing $O(n)$ ancestors of $T$
before we output the next solution;
traversing $O(1)$ ancestors is sufficient. 

%This would result in time delay of $O(n\alpha)$, where $\alpha$ denotes
%the time complexity required for a single run of
%{\sc Children}$(T,k)$. 
%To improve the delay to $O(\alpha)$, we employ, where  

\begin{algorithm}[t!]
  \caption{An algorithm to enumerate $\rho$-positive solutions in $\MS_k$ 
  for a given $k\in [0,q]$}
  % for $(V,\MC,I,\itemmap)$ and an item $i\in $I$ }
  \label{alg:enumalg}
  \begin{algorithmic}[1]
    \Require An instance $(V,\MC,I=[1,q],\itemmap)$, 
     and an item $k\in  [0,q]$
    \Ensure The set $\MS_k$ of   solutions to $(V,\MC,I,\itemmap)$ 
    \State Compute $\Max(V_{\angleb{k}})$; $d:=1$; 
    \label{line:enum_vk}
    \For{{\bf each} $T\in\Max(V_{\angleb{k}})$}
    \label{line:enum_for}
    \If{$k=\min \Item(T)$ (i.e., $T\in \MB_k$) and $\rho(T)>0$}
    \label{line:enum_if}
         \State  Output $T$; 
         \If {$k\in [1,q-1]$}
         \State {\sc Descendants}$(T,k,d+1)$
         \EndIf
    \EndIf
    \EndFor
    \label{line:enum_endfor}
  \end{algorithmic}
\end{algorithm}
  
\begin{algorithm}[t!]
  \caption{{\sc Descendants}$(T,k,d)$: Generating all $\rho$-positive descendant solutions}
  \label{alg:enumdesc}
  \begin{algorithmic}[1]
%   \Procedure{EnumDescendants}{$T,A_T,i_T$}\Comment{$A_T=\Item(T)$, $i_T=\min \Item(T)$}
\Require An instance $(V,\MC,I,\itemmap)$, $k\in   [1,q-1]$, 
% $\{\Max(V_{\angleb{i}}) \mid i\in  I\}$, 
a solution $T\in \MS_k$, 
%       where $k= \min \Item(T)$, 
the current depth $d$ of recursive call of {\sc Descendants}, and
a volume function $\rho:2^V\to\mathbb{R}$
\Ensure All $\rho$-positive descendant solutions of $T$ in $\MS_k$ 
 \For{{\bf each} item $j\in[k+1, q]\setminus \Item(T)$}
 \label{line:desc_for}
 \State Compute $\Max(T\cap V_{\angleb{j}})$;
 \label{line:desc_max}
    \For{{\bf each} component $S\in\Max(T\cap V_{\angleb{j}})$}
      \label{line:desc_inner_for}
      \If {$k= \min \Item(S)$,
               $j=\min\{i\mid i\in [k+1,q]\cap (\Item(S)\setminus \Item(T))\}$, \par
        \hskip\algorithmicindent $T= ${\sc Parent}$(S)$  (i.e., $S$ is a child of $T$), and $\rho(S)>0$  }
            \If{$d$ is odd}
              \State Output $S$
            \EndIf;   
            \State {\sc Descendants}$(S,k,d+1)$; 
            \If{$d$ is even}
                \State Output $S$
            \EndIf 
    \EndIf
   \EndFor
 \label{line:desc_inner_endfor}
 \EndFor   
 \label{line:desc_endfor}
  \end{algorithmic}
\end{algorithm}

\begin{lemma}   \label{lem:main_poly} 
Let $(V,\MC,I=[1,q],\sigma)$ be an instance. 
For each $k\in [0,q]$, all $\rho$-positive solutions in $\MS_k$ can be enumerated
in
$O\big( (n+q)q\delta(V_{\angleb{k}})\oraclex{1,t}+q\oraclex{2,t}+q\delta(V_{\angleb{k}})\oraclex{\rho,t}+(n^2+nq)q\delta(V_{\angleb{k}}) \big)$
 delay and  
$O\big(n(\oraclex{1,s}+\oraclex{2,s}+\oraclex{\rho,s}+n+q)\big)$
space.
\end{lemma} 
\begin{proof}
  Let $T\in\MS_k$ be a solution such that $\rho(T)\leq 0$. 
  In this case, $\rho(S)\le\rho(T)\leq 0$ holds
  for all descendants $S$ of $T$ since $S\subseteq T$.
  Then we do not need to make recursive calls for such $T$. 
  
  We analyze the time delay. 
  Let $\alpha$ denote the time complexity required for
  a single run of {\sc Children}$(T,k)$.
  By \lemref{child_candidate}(iii)
  and $\delta(T)\le\delta(V_{\angleb{k}})$,
  we have $\alpha=O\big((n+q)q\delta(V_{\angleb{k}})\oraclex{1,t}+q\oraclex{2,t}+(n^2+nq)q\delta(V_{\angleb{k}})\big)$. 
  In \algref{enumalg} and {\sc Descendants},
  we also need to compute $\rho(S)$ for all child candidates $S$.
  The complexity is $O(q\delta(V_{\angleb{k}})\theta_{\rho,\mathrm{t}})$
  since $\rho(S)$ is called at most $q\delta(V_{\angleb{k}})$ times.
  Hence we see that the time complexity of  \algref{enumalg} 
  and {\sc Descendants}
  without including recursive calls is $O(\alpha+q\delta(V_{\angleb{k}})\theta_{\rho,\mathrm{t}})$.

  During the execution of \algref{enumalg},
  {\sc Descendants}$(T,k,d)$
  may call {\sc Descendants$(T',k,d+1)$} recursively,
  where $T'$ is a child of $T$,
  or may go back to {\sc Descendants}$(T'',k,d-1)$
  by backtracking, where $T''$ is the parent of $T$. 
  We denote by $(T_1,d_1=1),(T_2,d_2),\dots,(T_m,d_m=1)$
  the sequence of $(T,d)$
  that are arranged in the visited order,
  where $m$ is the total number of all such $(T,d)$
  that equals to double the total times
  at which {\sc Descendants} is called plus one. 
  Note that a solution $T\in\MS$ may appear
  as $T_i$ for multiple $i\in[1,m]$;
  e.g., $T_1=T_m$ holds. 
  
  We show that, for any $i\in[1,m-2]$,
  at least one solution is output
  in {\sc Descendants}$(T_i,k,d_i)$,
  {\sc Descendants}$(T_{i+1},k,d_{i+1})$ and
  {\sc Descendants}$(T_{i+2},k,d_{i+2})$,
  by which we see that the delay for outputting a solution is
  $O(\alpha+q\delta(V_{\angleb{k}})\theta_{\rho,\mathrm{t}})$.

  \begin{itemize}
  \item 
  Suppose that $d_i$ is odd. If $d_{i+1}=d_i+1$,
  then $T_{i+1}$ is a child of $T_i$
  and should be output in {\sc Descendants}$(T_i,k,d_i)$.
  Otherwise (i.e., if $d_{i+1}=d_i-1$), 
  $T_{i+1}$ is the parent of $T_i$.
  Then $T_i$ should be output in {\sc Descendants}$(T_{i+1},k,d_{i+1})$. 
\item Suppose that $d_i$ is even.
  If $d_{i+1}=d_i+1$ and $d_{i+2}=d_{i+1}-1=d_i$,
  then it means that no recursive call is made
  in {\sc Descendants}$(T_{i+1},k,d_{i+1})$ and
  $T_i=T_{i+2}$.
  Then $T_{i+1}$ should be output in {\sc Descendants}$(T_{i+2},k,d_{i+2})$.
  If $d_{i+1}=d_i+1$ and $d_{i+2}=d_{i+1}+1$,
  then a child of $T_{i+1}$ is output in {\sc Descendants}$(T_{i+1},k,d_{i+1})$.
  The case of $d_{i+1}=d_i-1$ is analogous. 
  \end{itemize}
  
  \invis{
  From \algref{enumalg} and {\sc Descendants}, we observe
  \begin{itemize}
  \item[(i)] When $d$ is odd, the solution $S$ for any call {\sc Descendants}$(S,k,d+1)$ is output immediately before  {\sc Descendants}$(S,k,d+1)$ is executed; and
  \item[(ii)] When $d$ is even, the solution $S$ for any call
   {\sc Descendants}$(S,k,d+1)$
   is output  immediately after  {\sc Descendants}$(S,k,d+1)$ is executed.
  \end{itemize}
  Let $m$ denote the number of all calls of {\sc Descendants} during a whole
  execution of  \algref{enumalg}.
  Let $d_1=1,d_2,\ldots,d_m$ denote the sequence of depths $d$
  in each {\sc Descendants}$(S,k,d+1)$ of the $m$ calls.
  Note that $d=d_i$ satisfies (i) when $d_{i+1}$ is odd and $d_{i+1}=d_i+1$, 
  whereas $d=d_i$ satisfies (ii) when  $d_{i+1}$ is even and $d_{i+1}=d_i-1$.
  Therefore we easily see that during three consecutive calls 
  with depth $d_i$, $d_{i+1}$ and $d_{i+2}$,
  at least one solution will be output.
  This implies that the time delay for outputting a solution is
   $O(\alpha+q\delta(V_{\angleb{k}})\theta_{\rho,\mathrm{t}})$.  
  }

  We analyze the space complexity. 
  Observe that the number of calls  {\sc Descendants} whose executions
   are not finished  
  during an execution of \algref{enumalg} is the depth $d$
  of the current call {\sc Descendants}$(S,k,d+1)$. 
  In \algref{enumdesc}, 
  $|T|+d\leq n+1$ holds initially,
  and 
   {\sc Descendants}$(S,k,d+1)$ is called
  for a nonempty subset $S\subsetneq T$, where $|S|<|T|$.
  Hence $|S|+d\leq n+1$ holds when  {\sc Descendants}$(S,k,d+1)$ is called.
Then  \algref{enumalg} can be implemented to run 
 in $O(n(\beta+\theta_{\rho,\mathrm{s}}))$ space, where
  $\beta$ denotes the space required for a single run of {\sc Children}$(T,k)$.
  We have $\beta=O(\theta_{1,\mathrm{s}}+\theta_{2,\mathrm{s}}+n+q)$
  by \lemref{child_candidate}(ii).
  Then the overall space complexity is 
  $O\big(n(\theta_{1,\mathrm{s}}+\theta_{2,\mathrm{s}}+\theta_{\rho,\mathrm{s}}+n+q)\big)$.
  %\qed
\end{proof}

The volume function is introduced to impose a condition on the output solutions. 
For example, 
when $\rho(X)=|X|-p$ for a constant integer
$p$,
all solutions $X\in\MS_k$ with $|X|\ge p+1$ will be output.
In particular, all solutions in $\MS_k$ will be output for $p\le0$. 
In this case, we have 
$\theta_{\rho,\mathrm{t}}=\theta_{\rho,\mathrm{s}}=O(n)$,
and thus the delay is $O\big((n+q)q\delta(V_{\angleb{k}})\oraclex{1,t}+q\oraclex{2,t}+(n^2+nq)q\delta(V_{\angleb{k}})\big)$ and
the space is $O\big(n(\oraclex{1,s}+\oraclex{2,s}+n+q)\big)$. 

\thmref{main} is immediate from \lemref{main_poly}
since $\delta(V_{\angleb{k}})\le\delta(V)$ holds
by our assumption
that $\delta(Y)\le\delta(X)$ for subsets $Y\subseteq X\subseteq V$. 

\subsection{Enumerating Components}\label{sec:enum}

As we have done for a confluent system in \cite{HN.2022},
we can enumerate
all components in a given system  $(V,\MC)$ with $n=|V|\geq 1$
by using the algorithm proposed in the last subsection. 
For this, we construct an instance
$\mathcal{I}=(V,\MC,I=[1,n],\varphi)$
as follows.
Denote $V$ by $\{v_1,\dots,v_n\}$.
We set $I=[1,n]$
and define a function $\varphi:V\to 2^I$
to be $\varphi(v_k)\triangleq I\setminus\{k\}$ for each element $v_k\in V$.
For each subset $X\subseteq V$, let  $\Ind(X)$ denote the set of indices $i$ 
of elements $v_i\in X$; i.e.,   $\Ind(X)=\{i\in [1,n]\mid v_i\in X\}$,
and  $I_\varphi(X)\subseteq [1,n]$ 
denote the common item set over $\varphi(v)$, $v\in X$;
i.e.,   $I_\varphi(X) = \bigcap_{v\in X}\varphi(v)$.
Observe that $I_\varphi(X)=I\setminus\Ind(X)$. 

The following lemma is taken from \cite{HN.2022}.
In the original lemma, the given system is assumed to
be confluent, but the assumption can be dropped;
the confluency was not used anywhere in the proof.

\begin{lemma}[Lemma~9 in \cite{HN.2022}]
  \label{lem:compsol}
  Let $(V=\{v_1,\dots,v_n\},\MC)$ be a system with $n\geq 1$.
  The family $\MC$ of all components is equal 
  to the family $\MS$ of all solutions
  in the instance $(V,\MC,I=[1,n],\varphi)$. 
\end{lemma}
\begin{proof} Since any solution $S\in \MS$ is a component, it holds that 
  $\MC\supseteq \MS$.
  We prove that $\MC\subseteq \MS$.
  Let $X\in\MC$. 
  For any superset $Y\supsetneq X$,
  it holds that $I_{\varphi}(Y)=I\setminus \Ind(Y)\subsetneq I\setminus \Ind(X)=I_{\varphi}(X)$.
  The component $X$ is a solution in $(V,\MC,I,\varphi)$
  since no superset of $X$ has the same common item set as $X$. 
  %\qed
\end{proof}

Since the family  $\MC$ of components is equal to the family $\MS$ of solutions
to the instance $\mathcal{I}=(V,\MC,I,\varphi)$   by \lemref{compsol}, 
we can enumerate all components in $(V,\MC)$ by
running our algorithm on the instance $\mathcal{I}$. 
By $|I|=n$, we have the following corollary to \thmref{main}. 

\begin{corollary}\label{cor:comp} 
  Let $(V,\MC)$ be a system with $n=|V|\geq 1$
  and a volume function $\rho$.
  All $\rho$-positive components in $\MC$ can be enumerated 
  in $O\big(n^2\delta(V)\oraclex{1,t}+n\oraclex{2,t}+n\delta(V)\oraclex{\rho,t}+n^3\delta(V) \big)$ delay and 
  $O\big(n\oraclex{1,s}+n\oraclex{2,s}+n\oraclex{\rho,s}+n^2\big)$ space.
\end{corollary}

%%%%%%%%%%%%%%%%%%%%%%%%%%%%%%%%%%%%%%%%%%%%%%%%%%%%%%%%%%%%%%%%%%%%%%%%%%%%%%
\section{Concluding Remarks}
\label{sec:conc}

In this paper, we have shown that all solutions
in a given instance $(V,\MC,I,\sigma)$ can be enumerated
in polynomial-delay with respect to the input size;
an upper bound on the number of maximal components;
and the running times of the oracles
even when $(V,\MC)$ is an arbitrary system (\thmref{main}).
As a corollary to the theorem,
we have also shown that all components in $(V,\MC)$
are enumerable in polynomial-delay (\corref{comp}). 
The achievements generalize the result of \cite{HN.2022}
in which $(V,\MC)$ is restricted to a confluent system.

We have shown that, if maximal oracles are provided,
then we can enumerate all solutions 
in polynomial delay for any set system,
where a solution is defined in terms of items.
In other words, our item-based solutions can be enumerable in
polynomial delay
once maximal oracles that run in polynomial time
are available. 
In the meanwhile, as mentioned in \secref{prel},
it remains a challenging issue
to enumerate maximal subsets in polynomial-delay 
for various problems,
which is among our future work.

\bibliographystyle{elsarticle-num}
\bibliography{general}






% \end{thebibliography}

\end{document}